%Paper: gr-qc/9404006
%From: jbrown@unity.ncsu.edu
%Date: Wed, 6 Apr 1994 10:11:51 -0400 (EDT)

%% This is a plain TeX file.

\message{*** SIAM Plain TeX Proceedings Series macro package, version 1.0,
November 6, 1992.***}

\catcode`\@=11
\baselineskip=14truept

%%%  DIMENSIONS  %%%
\hsize=36truepc
\vsize=55truepc
\parindent=18truept
\def\firstpar{\parindent=0pt\global\everypar{\parindent=18truept}}
\parskip=0pt

%%%  FONTS  %%%
\font\tenrm=cmr10

\font\tensmc=cmcsc10

\font\ninerm=cmr9
\font\ninebf=cmbx9
\font\nineit=cmti9
\def\ninepoint{%
   \def\rm{\ninerm}\def\bf{\ninebf}%
   \def\it{\nineit}\baselineskip=11pt\rm}%

\fontdimen13\tensy=2.6pt
\fontdimen14\tensy=2.6pt
\fontdimen15\tensy=2.6pt
\fontdimen16\tensy=1.2pt
\fontdimen17\tensy=1.2pt
\fontdimen18\tensy=1.2pt

\font\ninerm=cmr9
\font\elevenrm=cmr10 scaled\magstephalf
\font\fourteenrm=cmr10 scaled\magstep 1
\font\eighteenrm=cmr10 scaled\magstep 3
\font\twelvebf=cmbx10 scaled\magstep 1
\font\elevenbf=cmbx10 scaled\magstephalf

\def\textfont{\elevenrm}

\def\headfont{\twelvebf}
\def\smallheadfont{\elevenbf}
\def\titlefont{\eighteenrm}

\def\authorfont{\fourteenrm}
\def\rheadfont{\tenrm}
\def\abstractfont{\tenrm}

%\font\eightsmc=cmcsc8

%% FOLLOWING LINE CANNOT BE BROKEN BEFORE 80 CHAR
\def\footnote#1{\baselineskip=11truept\edef\@sf{\spacefactor\the\spacefactor}#1\@sf
  \insert\footins\bgroup\ninepoint\hsize=36pc
  \interlinepenalty10000 \let\par=\endgraf
   \leftskip=0pt \rightskip=0pt
   \splittopskip=10pt plus 1pt minus 1pt \floatingpenalty=20000
\smallskip
\item{#1}\bgroup\baselineskip=10pt\strut
\aftergroup\@foot\let\next}
\skip\footins=12pt plus 2pt minus 4pt
\dimen\footins=36pc

%%%  CHAPTER OPENING MACROS  %%%
\def\startchapter{\topinsert\vglue54pt\endinsert}

\def\title#1\endtitle{\titlefont\centerline{#1}\vglue5pt}
%\vskip40truept\tenrm}

\def\lasttitle#1\endlasttitle{\titlefont\centerline{#1}
\vskip1.32truepc}
\def\author#1\endauthor{\authorfont\centerline{#1}\vglue8pt
\textfont}
\def\lastauthor#1\endlastauthor{\authorfont\centerline{#1}
\vglue2.56pc\textfont}

\def\abstract#1\endabstract{\baselineskip=12pt\leftskip=2.25pc
     \rightskip=2.25pc\abstractfont{#1}\textfont}

%%%  COUNTERS FOR HEADINGS  %%%
\newcount\headcount
\headcount=1
\newcount\seccount
\seccount=1
\newcount\subseccount
\subseccount=1
\def\secreset{\global\seccount=1}
 \def\subsecreset{\global\subseccount=1}

%%%  HEADINGS  %%%
\def\headone#1{\baselineskip=14pt\leftskip=0pt\rightskip=0pt
\vskip17truept\parindent=0pt
{\headfont\the\headcount\hskip14truept #1}
\par\nobreak\firstpar\global\advance\headcount by 0
   \global\advance\headcount by 1\secreset\vskip2truept\textfont}

\def\headtwo#1{\advance\headcount by -1%
   \vskip17truept\parindent=0pt{\headfont\the\headcount.%
   \the\seccount\hskip14truept #1}%\enspace\ignorespaces\firstpar
   \global\advance\headcount by 1\global\advance\seccount by 1
   \global\advance\subseccount by 1\subsecreset\vskip2pt\textfont}

 \def\headthree#1{\advance\headcount by -1\advance\seccount by -1
%   \advance\subseccount by -1%
   \vskip17truept\parindent=0pt{\smallheadfont\the\headcount.%
   \the\seccount.\the\subseccount\hskip11truept #1}\hskip6pt
   \ignorespaces
   \firstpar\global\advance\headcount by 1\global\advance
   \seccount by 1
   \global\advance\subseccount by 1\textfont}

%%%  REFERENCES  %%%
\newdimen\refindent@
\newdimen\refhangindent@
\newbox\refbox@
\setbox\refbox@=\hbox{\tenrm\baselineskip=11pt [00]}%   Default 2 digits
\refindent@=\wd\refbox@

\def\resetrefindent#1{%
	\setbox\refbox@=\hbox{\tenrm\baselineskip=11pt [#1]}%
	\refindent@=\wd\refbox@}

\def\Refs{%
	\unskip\vskip1pc
	\leftline{\noindent\headfont References}%
	\penalty10000
	\vskip4pt
	\penalty10000
	\refhangindent@=\refindent@
	\global\advance\refhangindent@ by .5em
        \global\everypar{\hangindent\refhangindent@}%
	\parindent=0pt\baselineskip=12pt\tenrm}

\def\sameauthor{\leavevmode\vbox to 1ex{\vskip 0pt plus 100pt
    \hbox to 2em{\leaders\hrule\hfil}\vskip 0pt plus 300pt}}

\def\ref#1\\#2\endref{\leavevmode\hbox to
\refindent@{\hfil[#1]}\enspace
#2\par}

%%%  OUTPUT  %%%
\def\rightheadline{\hfill\tensmc\rightrh\hskip2pc\tenrm\folio}
\def\leftheadline{\tenrm\folio\hskip2pc\tensmc\leftrh\hfill}

\global\footline={\hss\tenrm\folio\hss}% first page

\output{\plainoutput}
\def\plainoutput{\shipout\vbox{\makeheadline\pagebody\makefootline}%
  \advancepageno
  \ifnum\pageno>1
	\global\footline={\hfill}%
  \fi
  \ifodd\pageno
	\global\headline={\rightheadline}%
  \else
	\global\headline={\leftheadline}%
  \fi
  \ifnum\outputpenalty>-\@MM \else\dosupereject\fi}
\def\pagebody{\vbox to\vsize{\boxmaxdepth\maxdepth \pagecontents}}
\def\makeheadline{\vbox to\z@{\vskip-22.5\p@
  \line{\vbox to8.5\p@{}\rheadfont\the\headline}\vss}%
    \nointerlineskip}
\def\makefootline{\baselineskip24\p@\vskip-6\p@\line{\the\footline}}
\def\dosupereject{\ifnum\insertpenalties>\z@
% something is being held over
  \line{}\kern-\topskip\nobreak\vfill\supereject\fi}

\def\footnoterule{\vskip11pt\kern -4\p@\hrule width 3pc \kern 3.6\p@ }
%rule = .4 pt high

\catcode`\@=13

%%  BEGIN  %%
\def\leftrh{David Brown}
\def\rightrh{Black Hole Thermodynamics in a Box}
\startchapter
\lasttitle Black Hole Thermodynamics in a Box\endlasttitle

\lastauthor David Brown\footnote{$^*$}{Departments of Physics and
Mathematics,
North Carolina State University, Raleigh, NC 27695--8202}
\endlastauthor

\centerline{\bf Abstract}
\abstract Simple calculations indicate that the partition function
for a black hole is defined only if the temperature is fixed on a finite
boundary. Consequences of this result are discussed. \endabstract

\headone{The Black Hole Partition Function}
{}From the work of Gibbons and Hawking in the late 1970's came a very simple
prescription for the computation of the temperature of a static black hole
[1]:
\item{$\bullet$} Write the black hole metric in static coordinates;
\item{$\bullet$} Euclideanize ($t\to -it$) and periodically identify $t$;
\item{$\bullet$} Adjust the period to remove conical singularities.

\noindent The resulting period is the inverse temperature $\beta$. The
origin of
this prescription is a formal calculation of the partition function
$Z(\beta)$ as a functional integral over all Euclidean geometries $g$ with
period
$\beta$ and action $I[g]$. Some of the key features of this calculation
can be
captured in a `microsuperspace' version based on the metric ansatz [2]
$$ ds^2 = N^2(r) \, dt^2 + (1 - 2M/r)^{-1} dr^2 + r^2 d\Omega^2 \
.\leqno(1)$$
Let $t$ have the range $0\leq t< 2\pi$ and $r$ have the range $2M\leq r
<\infty$.
Also restrict $N$ so that $2\pi N(\infty) = \beta$ and $2\pi N(r) \sim
8\pi M
\sqrt{1 - 2M/r}$ near $r=2M$. The first restriction fixes the proper
period at infinty
to the inverse temperature. The second restriction insures that the metric
(1) describes a smooth geometry with no conical singularities, and with
topology
$R^2\times S^2$. The action
for (1) is a function of $M$ only, $I(M) = M\beta - 4\pi M^2$. A `toy'
partition
function can be constructed as the integral over $M$ of $\exp(-I(M))$. The
extremum of the action satisfies $0 = \partial I/\partial M = \beta-8\pi
M$.
(This is the classical equation of motion obtained by integrating
$(Nr/2)( 1 - 2M/r )^{-3/2} G^r_r$ over $t$ and $r$, where $G^r_r$ is the
$r$-$r$ component of the Einstein tensor.)
The solution for the extremum is a Euclidean black hole with $M
=\beta/(8\pi)$,
and the partition function is classically approximated by $\ln Z(\beta)
\approx -I(\beta/(8\pi)) = -\beta^2/(16\pi)$. The expectation value of
energy
is $\langle E\rangle \equiv -(\partial\ln Z /\partial\beta) \approx
\beta/(8\pi)$, which equals the extremal value of the mass parameter $M$.
An interpretation
of these results is that $Z(\beta)$ describes a system that contains a
black hole of mass $M$ and inverse temperature $\beta=8\pi M$.

What about pre--exponential factors in $Z(\beta)$? A simple calculation
shows that the second derivative $\partial^2 I/\partial M^2$ is negative
at the
extremum $M=\beta/(8\pi)$. Therefore, the extremum lies along a path of
steepest
{\it ascents}, not along a path of steepest descents. The Euclidean black
hole
does not dominate the integral for $Z(\beta)$, and should not be used to
approximate $Z(\beta)$. As a consequence, the conclusion
$\langle E\rangle = \beta/(8\pi)$ is unfounded. Formally, the Euclidean
black
hole makes an imaginary contribution to the partition function, and should
be
interpreted as an instanton that governs black hole nucleation [3].

So the prescription given above for the temperature of a black hole is not
justified, perhaps not even correct. Yet, it is tempting to believe in
that prescription because apparently it gives the correct result
$\beta= 8\pi M$ for the Hawking temperature. In order to understand this
puzzling situation,
consider a modified microsuperspace calculation [2]. As before, the metric
ansatz is given by Eq.~(1) with $0\leq t<2\pi$ and $2\pi N(r) \sim 8\pi M
\sqrt{1 - 2M/r}$ near $r=2M$. In this case, however, the system is placed
in a
finite `box' of size $R$ by restricting $r$ to the range $2M\leq r \leq R$
and fixing
the proper period at $r=R$ to the inverse temperature: $2\pi N(R) =
\beta$.
The action is $I(M) = R\beta (1 - \sqrt{1 - 2M/R}) - 4\pi M^2$, and a
toy partition function is constructed by integrating $\exp(-I(M))$ over
$M$.
The action is extremized for $M$ satisfying
$$\beta = 8\pi M\sqrt{1 - 2M/R} \ .\leqno(2)$$
(Again, this is related to the $G^r_r=0$ Einstein equation.)
There are two solutions to Eq.~(2), $M_1$ and $M_2$ with $M_1< M_2$.
$\partial^2 I/\partial M^2$
is negative at the extremum $M_1$, and $M_1\to\beta/(8\pi)$ as
$R\to\infty$ with
$\beta$ fixed. Thus, the Euclidean black hole with mass parameter $M_1$ is
an
instanton. On the other hand, $\partial^2 I/\partial M^2$
is positive at the extremum $M_2$, and $M_2\sim R/2 \to\infty$ as
$R\to\infty$ with $\beta$ fixed. It follows that the Euclidean black hole
with
mass parameter $M_2$ can be used for a steepest descents approximation to
$Z(\beta)$. In the classical approximation, $\ln Z(\beta) \approx -I(M_2)$
and the expectation value of energy is $\langle E\rangle \approx
R \bigl(1 - \sqrt{1 - 2M_2/R}\bigr)$.  $\langle E\rangle$ can be expanded
in powers
of $G M_2/R$ (where Newton's constant $G$ is set to unity) with the result
$ \langle E\rangle \approx M_2 + M_2^2/(2R)  +\cdots$. This
shows that the energy $\langle E\rangle$ inside the box $R$ equals
the energy at infinity $M_2$ {\it minus} the binding energy $-M_2^2/(2R)$
of a
shell of mass $M_2$ and radius $R$, which is the energy
associated with the gravitational field outside $R$ [4].

The calculation of $Z(\beta)$ above supports the conclusion that Eq.~(2)
gives the inverse
temperature at $R$ of a black hole with mass $M$. Note that the square
root
in (2) is the Tolman redshift factor for temperature in a stationary
gravitational field. In what sense can one say that the inverse
temperature
of a black hole is $8\pi M$? That statement does {\it not} follow from
taking
the limit $R\to\infty$ of Eq.~(2), since in that limit
$M\to\infty$ as well. Rather, $8\pi M$ is the inverse temperature obtained
by dropping
the Tolman redshift factor from Eq.~(2). Physically, this corresponds to
drilling
a small hole in the box and letting some radiation leak out to
infinity. The inverse temperature of the black hole as measured by the
radiation
at infinity is $8\pi M$.

It is important to recognize that the partition function, by itself, does
not
give the result $8\pi M$ (or any result) for the inverse temperature at
infinity.
With $\beta$ fixed at infinity, as in the first microsuperspace
calculation presented above, the (real part of the) partition function
does not exist.
This is a consequence of the physical fact that a gravitating system in an
infinitely
large cavity at nonzero temperature can {\it not} be in equilibrium,
because
black holes will form and grow without bound. In order to conclude that
the
inverse temperature at infinity is $8\pi M$, it is necessary
to supplement the partition function analysis with a further physical
argument.
In the argument given above, a small amount of radiation is allowed to
leak from
the box to infinity. Having argued in this way that the inverse
temperature
of an equilibrium black hole, as measured at infinity, is $8\pi M$, one
can
go a step farther and consider removing the box altogether. The black hole
will no longer be
in thermal equilibrium, but to the extent that it evolves relatively
slowly it is justified to identify $8\pi M$ as the inverse temperature at
infinity.
\vfill\eject

The calculation of the partition function for a system in a finite box
leads
to a corrected prescription for the temperature of a static black
hole:
\item{$\bullet$} Write the black hole metric in static coordinates;
\item{$\bullet$} Euclideanize ($t\to -it$) and periodically identify $t$;
\item{$\bullet$} Fix the proper period to $\beta$ at $R$;
\item{$\bullet$} Adjust the mass parameter $M$ to remove conical
singularities.

\noindent This yields two values for $M$. The larger $M$ is the mass
of a black hole with inverse temperature $\beta$ at
$R$. (This prescription holds as stated for 3+1 dimensional Einstein
gravity with a negative or vanishing cosmological constant [5]. In other
cases there might be more or fewer than two extrema $M$. In order to
distinguish among the instantons and the stable or quasi--stable black
holes, the sign of $\partial^2 I/\partial M^2$ must be checked for each
extremum.)

\headone{Temperature of Gravitating Systems}
For the canonical partition function $Z(\beta)$ of a gravitating system
the
temperature must be fixed at a finite boundary $B$. This has an important
consequence: Since temperature redshifts and blueshifts in stationary
gravitational fields, one must allow  the
temperature to be fixed to different values at different points on $B$.
In other words, gravitating systems are not characterized by a single
temperature
but instead by a temperature {\it field} on the boundary of the system
[6].
Correspondingly, the partition function is a functional $Z[\beta]$
of the inverse temperature field on $B$. For a typical problem
(such as the microsuperspace calculation of the previous section), it is
possible
to choose the temperature to be a constant on
$B$, in which case $B$ coincides with an isothermal surface for the
system. However, experience with the Kerr black hole shows that this must
be
viewed as a particular choice of boundary conditions, not the most general
choice.
What happens in the Kerr case is that the angular velocity of
the black hole with respect to observers who are at rest in the stationary
time slices enters as a chemical potential conjugate to angular momentum.
It
turns out that the constant temperature surfaces and the constant angular
velocity surfaces do not coincide. Therefore it is necessary to allow for
{\it some}
thermodynamical data, either the temperature or the chemical potential or
both,
to vary across the boundary. This conclusion might seem disturbing at
first,
since traditionally one of the purposes of thermodynamics has been
to provide a characterization of systems in terms of only a few
parameters. Nevertheless, the thermodynamical formalism that results from
a
generalization to non--constant thermodynamical data
has a number of compelling features. In particular, thermodynamical data
is brought into direct correspondence with canonical boundary data, and in
the
process an intimate connection between thermodynamics and dynamics is
revealed [6].

\Refs

\ref 1\\G.~W. Gibbons and S.~W. Hawking, {\it Phys. Rev.} {\bf 15} (1977)
2752.\endref

\ref 2\\This is based on B.~F. Whiting and J.~W. York,
{\it Phys. Rev. Lett.} {\bf 61} (1988) 1336; H.~W. Braden, J.~D. Brown,
B.~F. Whiting, and J.~W. York, {\it Phys. Rev.} {\bf D42} (1990)
3376.\endref

\ref 3\\D.~J. Gross, M.~J. Perry, and L.~G. Yaffe, {\it Phys. Rev.} {\bf
D25}
(1982) 330; J.~W. York, {\it Phys. Rev.} {\bf D33} (1986) 2092.\endref

\ref 4\\J.~D. Brown and J.~W. York, {\it Phys. Rev.} {\bf D47} (1993)
1407.\endref

\ref 5\\J.~D. Brown, J.~Creighton, and R.~B. Mann, submitted to
{\it Phys. Rev.} {\bf D}.\endref

\ref 6\\J.~D. Brown, E.~A. Martinez, and J.~W. York, {\it Phys. Rev.
Lett.}
{\bf 66} (1991) 2281; J.~D. Brown and J.~W. York, {\it Phys. Rev.} {\bf
D47}
(1993) 1420; J.~D. Brown and J.~W. York, in {\it Physical Origins of Time
Asymmetry}, edited by J.~J. Halliwell, J.~Perez--Mercader, and W.~Zurek
(Cambridge University Press, Cambridge, 1994).\endref

\bye